\documentclass[runningheads]{llncs}
\usepackage{amsmath}
\usepackage[ruled,vlined]{algorithm2e}
\usepackage{bm}
\usepackage{comment}
\usepackage{booktabs}
\usepackage{diagbox}
\usepackage{float}
\usepackage{multirow}
\usepackage[T1]{fontenc}
\usepackage[misc,geometry]{ifsym}
\usepackage{graphicx}
\usepackage{xcolor}
\begin{document}
\title{Improved 3D Whole Heart Geometry from Sparse CMR Slices}
%
%
\author{Yiyang Xu\inst{1}\textsuperscript{(\Letter)} \and
Hao Xu\inst{1,4} \and
Matthew Sinclair\inst{2,5} \and
Esther Puyol-Ant\'on\inst{1,5}  \and
Steven A Niederer \inst{2} \and
Amedeo Chiribiri \inst{1} \and
Steven E Williams \inst{1,3} \and
Michelle C Williams \inst{3} \and
Alistair A Young \inst{1} }
\authorrunning{Y. Xu et al.}
%
\institute{School of Biomedical Engineering and Imaging Sciences, King’s College
London, London, UK \\
\email{yiyang.1.xu@kcl.ac.uk} \and
Imperial College London, London, UK \and
University/BHF Centre for Cardiovascular Science, University of Edinburgh, Edinburgh, UK \and
College of Mathematical Medicine, Zhejiang Normal University, Zhejiang, China \and
HeartFlow, Inc., Mountain View, USA
}
\maketitle              
\begin{abstract}
Cardiac magnetic resonance (CMR) imaging and computed tomography (CT) are two common non-invasive imaging methods for assessing patients with cardiovascular disease. CMR typically acquires multiple sparse 2D slices, with unavoidable respiratory motion artefacts between slices, whereas CT acquires isotropic dense data but uses ionising radiation. In this study, we explore the combination of Slice Shifting Algorithm (SSA), Spatial Transformer Network (STN), and Label Transformer Network (LTN) to:  1) correct respiratory motion between segmented slices, and 2) transform sparse segmentation data into dense segmentation. All combinations were validated using synthetic motion-corrupted CMR slice segmentation generated from CT in 1699 cases, where the dense CT serves as the ground truth. In 199 testing cases,  SSA-LTN achieved the best results for Dice score and Huasdorff distance ($ 94.0\%$ and 4.7 mm respectively, average over 5 labels) but gave topological errors in 8 cases. STN was effective as a plug-in tool for correcting all topological errors with minimal impact on overall performance ($ 93.5\%$ and 5.0 mm respectively). SSA also proves to be a valuable plug-in tool, enhancing performance over both STN-based and LTN-based models. The code for these different combinations is available at \url{https://github.com/XESchong/STACOM2024}.
 \and Image Registration
\end{abstract}
\section{Introduction}
Accurate cardiac 3D geometry is required for evaluation of cardiovascular disease, the leading cause of death worldwide~\cite{uk_cad,WHO}. Reconstruction of accurate cardiac 3D geometry enables more precise and accurate evaluation of disease than current clinical metrics~\cite{Mauger2023l,marica}. Two common non-invasive imaging techniques for determining treatment are coronary computed tomography (CT) angiography~\cite{scotheart} and cardiac magnetic resonance (CMR) imaging~\cite{cmr}. CT captures dense geometry information at high isotropic resolution (0.5mm) but requires ionizing radiation. CMR is non-ionizing but is typically performed in sparse 2D slices, often in short axis (10mm apart) and long axis orientations, requiring multiple breath-holds and introducing motion artefacts between slices~\cite{motion_why} due to differences in patient breath-hold position. Two promising methods for 3D geometry reconstruction are label transformer networks (LTN) for shape reconstruction from sparse views~\cite{10.1007/978-3-031-35302-4_26} and spatial transformer networks (STN) for atlas registration~\cite{sinclair2020atlasistn}. However, an extensive evaluation of the performance of these methods, separately and combined, has not been performed. Here, we extensively evaluate the combination of slice shifting algorithm (SSA) for motion correction, LTN and STN modules for reconstruction of 3D geometry from CMR slice data. For the STN module, we evaluate two methods, one which registers an atlas to dense segmentation data (DSTN) and the other registers the atlas to sparse segmentation data (SSTN). We generate simulated CMR sparse segmentation data from a large CT cohort~\cite{scotheart} so that all methods can be compared using the same ground truth. 
\paragraph{Related work} 
Methods to correct motion artefact in CMR data include Chandler et al.~\cite{motion_correction_m4}, a slice-to-volume registration approach employing rigid transformations only. Another approach proposed by Sinclair et al.~\cite{mattmotion} utilizes a segmentation-based method, where segmentations are iteratively registered between slices to determine the in-plane shift.  Additionally, there is a growing interest in leveraging artificial intelligence (AI) models to either explicitly or implicitly eliminate motion artefacts across various organs~\cite{mc_ai_2,motion_correction_AI_one}. Geometry reconstruction involves transforming a sparse label map into a dense label map~\cite{10.1007/978-3-031-35302-4_26}. Wang et al.~\cite{wang2021joint} proposed an iterative latent optimisation network that simultaneously addresses super resolution and motion correction tasks for cardiac segmentation results, which demonstrates good performance over two CMR datasets. Besides that, Chen et al.~\cite{CHEN2021102228} presented MR-Net, a mesh construction method relying on sparse information from point clouds. 
Furthermore, Kong et al.~\cite{Kong_2021} incorporated a template mesh into a deep learning-based model for whole heart mesh reconstruction, leveraging prior knowledge of the surface topology.

\paragraph{Contributions}
Our main contributions can be summarized as follows:
\begin{itemize}
    \item We examine six combinations of SSA, LTN and STN modules to provide accurate and topologically correct reconstructions. 
    \item We show that STN can serve as a plug-in module for addressing topological errors with minimal impact on overall performance. 
    \item We provide an efficient implementation of SSA using \textit{PyTorch} and show enhanced performance when combined with both SSTN-based and LTN-based models.
   \item We show that LTN-DSTN demonstrated increased robustness in scenarios where 3-chamber long axis information is missing, a common occurrence in real CMR data acquisition. 
\end{itemize}
\section{Methodology}
\begin{figure}
    \centering
    \includegraphics[scale=0.17]{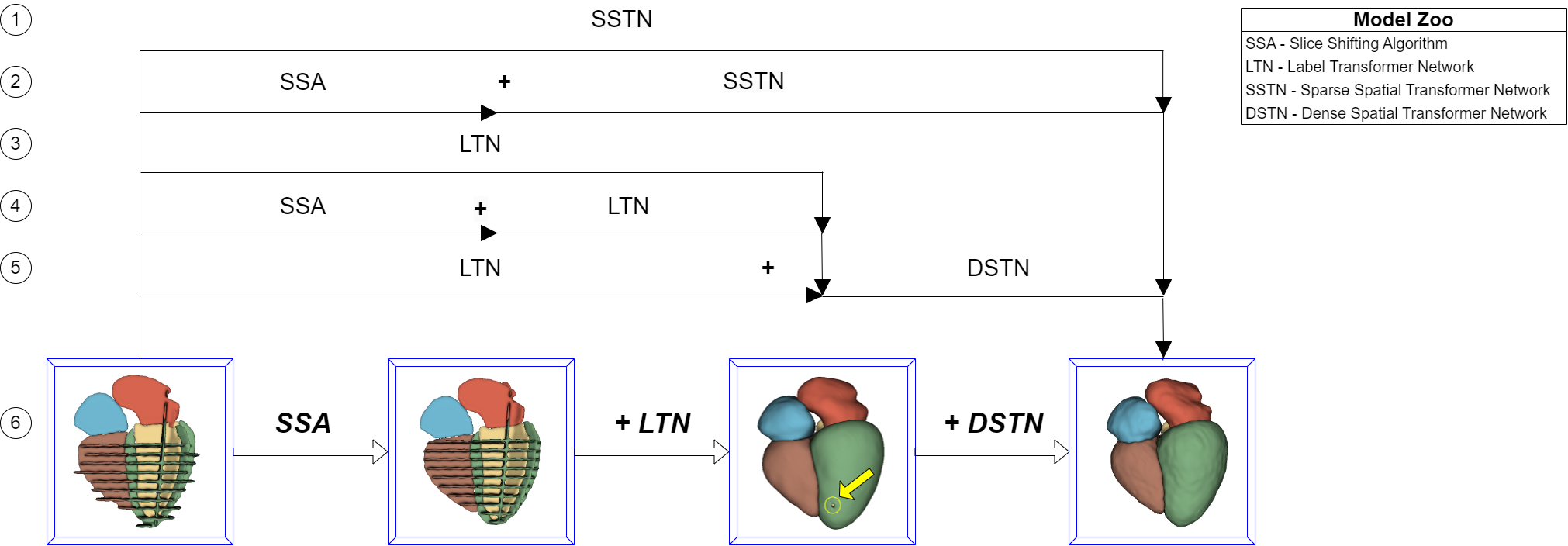}
    \caption{Overview of experiments: six combinations of SSA, LTN and STN (sparse and dense versions) were examined. The yellow circle denotes a topological error from LTN, corrected by STN (zoom in for better view).}
    \label{model zoo}
\end{figure}
\subsubsection{Slice Shifting Algorithm (SSA)} For the motion correction task, we adapted an iterative registration algorithm~\cite{mattmotion}. During each iteration, the label map undergoes adjustments aimed at optimizing its alignment with the overlapping regions of other slices. As in~\cite{mattmotion}, adjustments are conducted solely with in-plane 2D translation. Details are provided in the supplementary section. We also provide code for an efficient implementation of this method in \textit{PyTorch}.
\subsubsection{Spatial Transformer Network (STN)}
Derived from the Atlas-ISTN network~\cite{sinclair2020atlasistn}, our approach involves the application of the spatial transformer network (STN) and transformation computation module as the core model for addressing the geometry reconstruction task in CMR data. The primary network architecture is similar to the registration component of the Atlas-ISTN framework~\cite{sinclair2020atlasistn}. However, the input label map now could be a sparse data volume obtained from the stack of short axis (SAX) and long axis (LAX) views, which is denoted as $\hat{y}_{i}$. Another input for STN is a dense atlas label map $y^{a}$, which can be derived from various sources: 1) mean of training data 2) sample data from training dataset 3) publicly available cardiac atlas.  Mathematically, it can be represented as:
\begin{equation}
    \{v_{i}, T_{i}\} = \bm{D}_{\theta_{d}}(\hat{y}_{i}, y^{a})
\end{equation}
where $\theta_{d}$ are the parameters for the STN. After passing through the transformation computation module $\bm{C}$, the affine transformation $T_{i}$  and the stationary velocity field (SVF) $v_{i}$ undergo additional processing to generate the final deformation fields denoted as $\Phi_{i}$ and $\Phi_{i}^{-1}$. Specifically, $\Phi_{i}$ is the forward transformation from patient space to atlas space and $\Phi_{i}^{-1}$ is the inverse transformation from atlas space to the patient space.
\begin{figure}
    \centering
    \includegraphics[scale = 0.4]{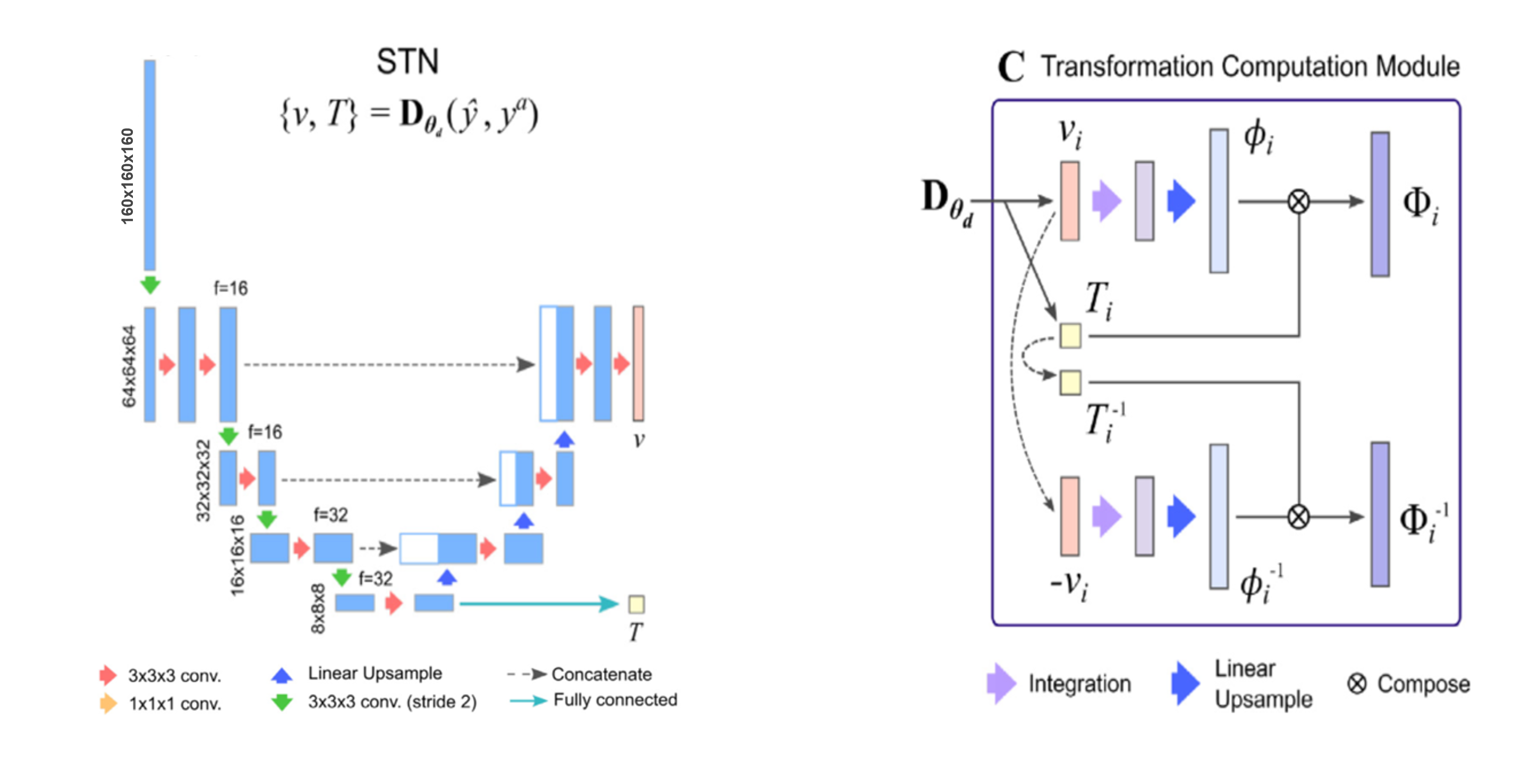}
    \caption{STN architecture (left) and  Transformation Computation Module (right)~\cite{sinclair2020atlasistn}}
    \label{stn_c}
\end{figure}
Finally, the forward pass of the STN framework with the input sparse volume $\hat{y}_{i}$ and atlas label map $y^{a}$ is given as:
\begin{equation}
     \{\Phi_{i}, \Phi_{i}^{-1}\} = \bm{C}(\bm{D}_{\theta_{d}}(\hat{y}_{i}, y^{a}))
\end{equation}
where $i$ is the index for the sample in the patient space. Given the dense ground truth label map volume $y^{dense}_{i}$ for the CMR data, the STN continues to utilize the previous atlas-to-segmentation loss $L_{a2s}$ and segmentation-to-atlas loss $L_{s2a}$, denoted as $L^{dense}_{a2s}$ and $L^{dense}_{s2a}$in the context of CMR data.
\begin{equation}
    L^{dense}_{a2s} = \sum_{i}\sum_{j=2}^{c}||y^{dense}_{i,j}-{y}_{j}^{a}\circ \Phi_{i}^{-1}||^{2}
    \label{L_a2s}
\end{equation}
\begin{equation}
    L^{dense}_{s2a} = \sum_{i}\sum_{j=2}^{c}||y_{j}^{a} - y^{dense}_{i,j}\circ \Phi_{i}||^{2}
    \label{L_s2a}
\end{equation}
where $j$ is the index for each channel of CMR data and $c$ is the total number of channels. Additionally, a regularisation term is introduced to control the smoothness of the predicted non-rigid displacement fields ${u}_{i}$, i.e.,
\begin{equation}
     L_{reg} = \sum_{i}||\Delta u_{i}||^{2}
\end{equation}
\begin{equation}
    \phi = I+u
\end{equation}
where $\phi$ in the transformation computation module is represented as a combination of two components, i.e., identity transformation $I$ and non-rigid displacement field $u$. Hence, the total loss is defined as:
\begin{equation}
     L =  L_{a2s} + L_{s2a} + \lambda L_{reg}
\end{equation}
where $\lambda$ is the weight for the regularisation term. A larger value of the hyperparameter $\lambda$ imposes a greater penalty on the non-rigid displacement field $u$, which tends to encourage the learning of a more rigid transformation within the STN framework. Based on the input for the training of STN, we denoted SSTN for the sparse volume input and DSTN for the dense volume input respectively.
\subsubsection{Label Transformer Network (LTN)} 
Drawing inspiration from the LC-U-Net~\cite{10.1007/978-3-031-35302-4_26}, we developed a 3D U-Net to perform the geometry reconstruction task, referred to as LTN in the subsequent sections. LTN is intended to learn the mapping between the sparse volume extracted from CMR data and their corresponding dense ground truth. During training, the mean square error (MSE) loss is employed to compute the total training loss, which compares the predictions $y'$ from LTN with the dense ground truth $y$. Mathematically, it is computed as:
\begin{equation}
    L = \sum_{i}\sum_{j}||y'_{i,j}-{y}_{i,j}||^{2}
    \label{L_ltn}
\end{equation}
where $i$ is the index for CMR data and $j$ is the index for each label of CMR data.
\section{Experiments}
\subsubsection{Synthetic CMR dataset} Based on the segmentation of SCOT-HEART CT data~\cite{scotheart,xu2021heart}, a cardiac coordinate system is established by identifying specific anatomical landmarks, such as the mitral valve, tricuspid valve and apex~\cite{10.1007/978-3-031-35302-4_26}. After this, we define the location of the planes corresponding to the short and long axis for the synthetic CMR data. The CT data, originally in the patient coordinate space, is then resampled into the cardiac coordinate space. Using the location information of each plane, we generate multiple synthetic CMR slices from CT data. These synthetic slices include random shifts for slice positions to replicate CMR slice planning errors.
Furthermore, we simulate respiratory motion artefacts by implementing a uniform random distribution for in-plane shifts. The main benefit of using synthetic CMR data is that we have paired motion-corrupted sparse data and dense ground truth volume for supervised-style training. Finally, the split of training, validation, and testing datasets in the synthetic CMR dataset is 1400, 100 and 199 cases, respectively.
\subsubsection{Implementation details} For SSA, each iteration begins by processing the long axis (LAX) slice of CMR data, with a maximum of 5 iterations in total. Usually, convergence is reached after approximately 3 iterations with our synthetic CMR data. SSA is implemented in \textit{PyTorch}, achieving a processing speed five times faster than the original implementation \cite{mattmotion}. For the STN, the input size is [160, 160, 160], comprising one background channel and five foreground labels: left ventricle myocardium (LVM), left ventricle blood pool (LV), left atrium (LA), right ventricle blood pool (RV) and right atrium (RA). To check the topological errors, the mesh for each channel is generated by using the Marching Cubes algorithm \cite{marchcube}. The learning rate and lambda values are set to $0.001$ and $2000$, respectively. The initial atlas map is selected from an example case in the training dataset. 
Unlike Atlas-ISTN, the update procedure for the atlas label map is disabled during training.
For the upcoming discussion, the model zoo is primarily categorized into two main groups: SSTN-based and LTN-based. Specifically, this includes SSA-SSTN and SSTN under the first group, and LTN, LTN-DSTN, SSA-LTN and SSA-LTN-DSTN under LTN-based models. All trainable models are trained for 50 epochs using the \textit{Nvidia RTX A500} GPU.
\subsubsection{Evaluation metrics} We utilize the Dice score~\cite{dice_score} and Hausdorff distance (HD)~\cite{hausdorff} to assess the similarity between the predicted label map and the ground truth. To quantify predictions in terms of topological accuracy, we employ the Euler characteristic to assess whether the predicted structure adheres to the topological constraints of heart. Specifically, the predicted mesh of each channel should satisfy the following formula to be considered topologically plausible \cite{euler}:
\begin{equation}
    V - E + F = 2
\end{equation}
where $V$, $E$, and $F$ denote the number of vertices, edges, and faces of the predicted mesh, respectively. 
\subsubsection{Results}
\begin{table}[h]
\begin{center}
\renewcommand\arraystretch{1.3}
\scalebox{0.7}{\begin{tabular}{|c|c|c|c|c|c|c|}
 \hline
 Models    & \backslashbox{Metric}{Label}    & LVM  & LV    & RV  & RA  & LA\\
  \hline
  \multirow{2}*{\textcircled{1}SSTN}
                    & Dice$(\%)$ $\uparrow$    & $85.58\pm3.60$ & $93.18\pm1.66$ & $91.31\pm2.68$ & $87.10\pm4.43$ & $86.91\pm4.18$ \\
 \cline{2-7}
    & HD$(mm)$ $\downarrow$    &$5.94\pm1.79$ & $5.31\pm 1.26$ & $7.72\pm2.90$ & $8.36\pm2.91$ & $7.84\pm2.14$ \\
  \hline
  \multirow{2}*{$\textcircled{2}\text{SSA-SSTN}^{\dagger}$}
                    & Dice$(\%)$ $\uparrow$    & $89.64\pm3.98$ & $94.87\pm1.80$ & $94.02\pm2.63$ & $88.97\pm3.90$ & $90.14\pm3.32$ \\
 \cline{2-7}
    & HD$(mm)$ $\downarrow$    &$5.08\pm1.89$ & $4.65\pm 1.45$ & $6.45\pm2.90$ & $8.13\pm3.15$ & $6.96\pm2.25$ \\
\hline
\hline
  \multirow{2}*{$\textcircled{3}\text{LTN}^{\ddagger}$}
                    & Dice$(\%)$ $\uparrow$    & $90.10\pm3.33$ & $95.35\pm1.39$ & $94.85\pm1.98$ & $91.10\pm2.87$ & $92.09\pm2.26$ \\
 \cline{2-7}
    & HD$(mm)$ $\downarrow$    &$4.59\pm4.08$ & $3.97\pm 1.11$ & $4.84\pm5.30$ & $7.74\pm8.58$ & $5.71\pm1.66$ \\
\hline
  \multirow{2}*{\textcircled{4}\textbf{SSA-LTN}}
                    & Dice$(\%)$ $\uparrow$    & $92.00\pm4.0$  & $96.25\pm1.70$    & $96.09\pm2.19$  & $92.13\pm2.52$    & $93.39\pm2.02$ \\
 \cline{2-7}
    & HD$(mm)$ $\downarrow$    & $3.78\pm2.21$   & $3.61\pm3.05$   & $4.34\pm7.05$ & $6.16\pm2.27$    & $5.53\pm2.00$ \\
\hline
  \multirow{2}*{$\textcircled{5}\text{LTN-DSTN}^{\ddagger}$}
                    & Dice$(\%)$ $\uparrow$    & $89.60\pm3.36$  & $95.06\pm1.38$    & $94.62\pm1.92$  & $91.00\pm2.81$    & $91.75\pm2.39$ \\
 \cline{2-7}
    & HD$(mm)$ $\downarrow$    & $4.59\pm1.56$   & $4.25\pm1.02$   & $5.33\pm2.23$ & $6.55\pm2.55$    & $5.91\pm2.16$ \\
\hline
  \multirow{2}*{$\textcircled{6}\text{SSA-LTN-DSTN}^{\ddagger}$}
                    & Dice$(\%)$ $\uparrow$    & $91.29\pm3.94$  & $95.87\pm1.68$    & $95.65\pm2.16$  & $91.75\pm2.58$    & $92.92\pm2.07$ \\
 \cline{2-7}
    & HD$(mm)$ $\downarrow$    & $4.26\pm1.71$   & $3.85\pm1.13$   & $4.79\pm2.06$ & $6.36\pm2.57$    & $5.65\pm1.97$ \\
\hline
\end{tabular}}
    \end{center}
    \caption{Performance of various models across different foreground channels of synthetic CMR dataset: each entry is in $mean \pm std$ format, where the best model is highlighted in bold. $\dagger$ denotes $p< 0.001$ for the difference of the Dice score between SSA-SSTN and SSTN. $\ddagger$ denotes $p< 0.001$ for the difference of the Dice score between current model and SSA-LTN.}
    \label{tab1}
\end{table}
In table \ref{tab1}, we present the performance for the geometry reconstruction task and motion correction task across various models. A comparison between SSTN and SSA-SSTN reveals that SSA-SSTN consistently achieves a higher Dice score and lower Hausdorff distance across all foreground channels. This underscores the efficacy of separately addressing motion correction and geometry reconstruction tasks, as opposed to concurrently addressing both tasks, where SSA is mainly for the motion correction task and SSTN is primarily for the geometry reconstruction task. We can also observe the advantage of addressing two tasks separately by examining the LTN-based model. This also indicates that SSA can serve as a versatile tool for enhancing the capabilities of any models. Furthermore, by integrating a pre-trained registration-based model (DSTN) after the dense volume prediction from LTN, we observed a slight performance decrease compared to using LTN alone. Although statistically significant, we believe that a drop of  $<1\%$ Dice and $<1$ mm Hausdorff distance will have minimal impact on resulting 3D shape statistics. However, DSTN proved beneficial in reducing cases of topological errors which has a big impact on computational shape analyses. Initially, 8 cases in the entire testing dataset exhibited topological errors based on the Euler characteristic for predictions from LTN. After applying DSTN, all these topological error cases were successfully rectified. Comparing the best models from SSTN-based and LTN-based approaches, SSA-LTN achieved the highest performance based on the Dice score and Hausdorff distance across all channels. Notably, the LTN-based model demonstrated superior performance in the atrium regions (LA and RA).
\begin{figure}[h]
    \centering
    \includegraphics[scale = 0.31]{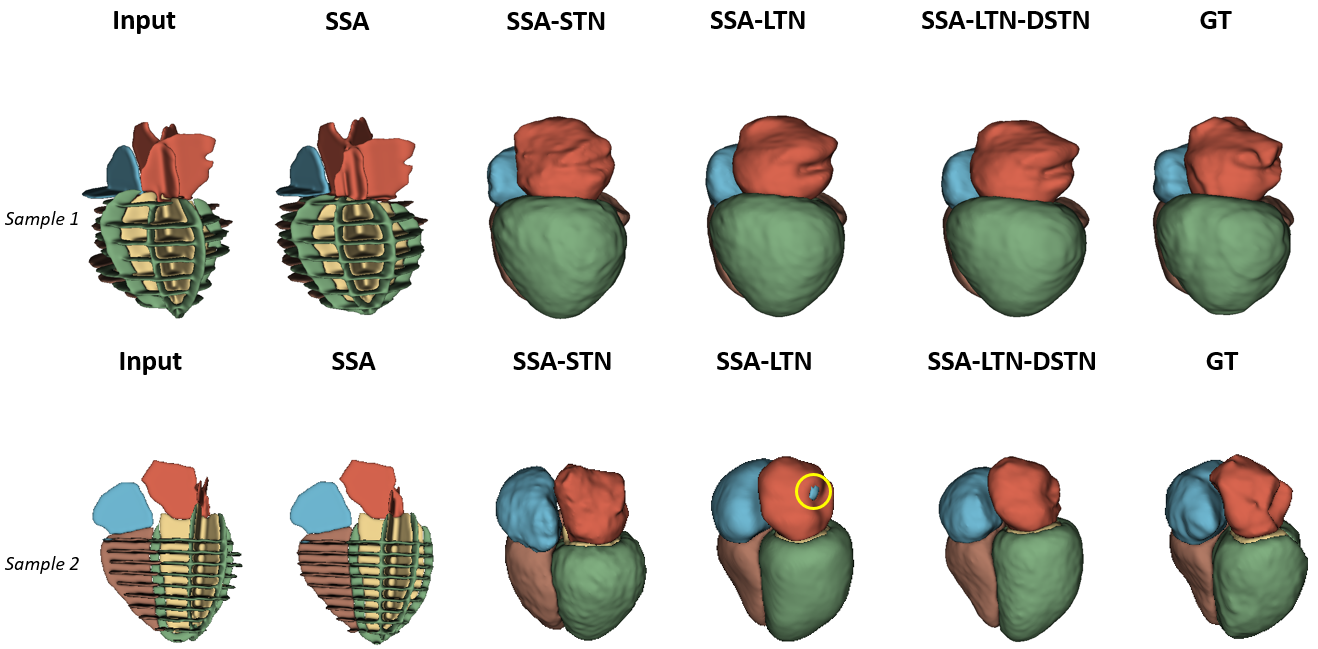}
    \caption{Visualisation of the performance for SSA-STN, SSA-LTN and SSA-LTN-DSTN, where GT is ground truth dense volume from CT data.}
    \label{fig 3}
\end{figure}

Based on the visual results shown in figure \ref{fig 3}, the first row demonstrates that SSA-STN's predictions exhibit finer details compared to those produced by SSA-LTN. However, SSA-LTN's predictions closely resemble the ground truth shape of the testing data, presenting a smoother surface. Importantly, the SSA component effectively reduces motion artefacts in synthetic CMR data. In the second row of figure \ref{fig 3}, a topological error is evident in SSA-LTN's prediction, specifically in the LA part. In contrast, SSA-LTN-DSTN, which incorporates atlas information during inference, successfully corrects the topological error and ensures topological plausibility. 
\begin{table}[h]
\centering
\scalebox{0.9}{
\begin{tabular}{l | c |c c |c c | c c | c c | c l}
\hline
\multicolumn{2}{c}{Labels} &  \multicolumn{2}{c}{LVM} & \multicolumn{2}{c}{LV} &  \multicolumn{2}{c}{RV} & \multicolumn{2}{c}{RA} &\multicolumn{2}{c}{LA}  \\
\hline
   Model  & Testing dataset & Dice & \text{ }\text{ }$\Delta$ & Dice & \text{ }\text{ }$\Delta$ & Dice & \text{ }\text{ }$\Delta$ & Dice & \text{ }\text{ }$\Delta$ & Dice & \text{ }\text{ }$\Delta$ \\

\hline
     
\multirow{3}*{SSTN} & SAX\&2/3/4 &  85.58  &        & 93.18  &       & 91.31 &       &  87.10 &       &  86.91  &    \\
 \cline{2-12}
& SAX\&2/4   &  83.87  & $1.71$  & 92.11  & $1.07$ & 90.80 & $0.51$  & 85.42  & $1.68$  & 83.77  & $\bm{3.14}$   \\
 \cline{2-12}
& SAX \& 4   &80.87   &  $\bm{4.71}$ &90.59   &  $2.59$ & 90.10& $\bm{1.21}$  & 83.84   & $3.26$  & 76.67 & $\bm{10.24}$\\
\hline

\multirow{3}*{LTN-DSTN}

&SAX\&2/3/4 &  89.60  &                       & 95.06  &                          & 94.62 &                          & 91.00 &                          & 91.75   &    \\
 \cline{2-12}
&SAX\&2/4   & 88.78   & $\bm{0.82}$  &94.73   & $\bm{0.33}$  &94.11  & 0.51  &89.76   & $\bm{1.24}$ &88.50   & 3.25   \\
 \cline{2-12}
&SAX \& 4   &83.94   &  5.66  &92.76   &  $\bm{2.30}$ &93.06 &  1.56 &88.25    &$\bm{2.75}$  &79.35 &12.40\\
\hline
\end{tabular}}
\caption{Robustness assessment between SSTN and LTN-DSTN, where $\Delta$ is the difference between the current row and the first row for a specific channel and bolded values signify a less substantial decrease.}
\label{table 2}
\end{table}

Furthermore, to assess the models' generalisation ability, we modified the testing dataset by reducing the number of slices. Initially, we omitted one long axis (LAX) slice, specifically the 3-chamber view. Subsequently, we removed another LAX slice, the 2-chamber view, leaving only a single 4-chamber LAX slice in the testing data. We retained the 4-chamber view as it provides crucial information about the right atrium. We denoted the testing data obtained by the above procedures as SAX\&2/3/4, SAX\&2/4, and SAX\&4, respectively. Evaluation from table \ref{table 2} indicates a consistent decline in the performance of SSTN and LTN-DSTN as the amount of information in the testing data decreases. However, when comparing the performance of each model between the first row and second row of table \ref{table 2}, the decline in the performance of LTN-DSTN was less pronounced compared to SSTN across most channels. This trend was not observed when considering the testing data containing only the sole 4-chamber LAX slice, where LTN-DSTN showed a notable decrease in performance for the LA channel. 
\section{Conclusion and Discussion}
We present an extensive evaluation of the combination of SSA, LTN and STN modules for the first time. Rather than simultaneously addressing motion correction and geometry reconstruction tasks, as in SSTN or LTN separately, our findings indicate that a sequential modular approach yields superior performance. SSA-LTN provides the best Dice, but also results in topological errors. An important advantage of STN is its ability to ensure the topological plausibility of the predicted heart shape, achieved through the incorporation of an atlas label map during both training and testing phases. Moreover, it can be seamlessly integrated as a plug-in tool for any models. As observed in the LTN-based model, it effectively corrected topological errors with minimal impact on overall performance. Furthermore, LTN-DSTN demonstrates increased robustness in scenarios where 3-chamber LAX information is missing, a common occurrence in real CMR data acquisition. Additionally, SSA serves as a valuable plug-in tool, which is validated for enhanced performance over both the SSTN-based and LTN-based models. Limitations of our work include reliance on segmentations, which poses challenges if the segmentation process yields inaccurate results. Moreover, as our motion correction task only accounts for in-plane shifting, it may alter the position of the heart within the 3D patient coordinate space after passing through SSA. In future endeavours, we plan to explore the integration of both rigid and non-rigid shifting techniques to effectively handle motion correction tasks. Also, we are using an example case as the atlas for STN training and testing. In the future, we could explore and apply different sources of atlas. Additionally, we aim to apply the combined modules to real-world datasets for further validation and practical implementation.
\begin{credits}
\subsubsection{\ackname}
We would like to acknowledge funding from the EPSRC Centre for Doctoral Training in Smart Medical Imaging (EP/S022104/1) and the Wellcome/EPSRC Centre for Medical Engineering (WT203148/Z/16/Z).
We also acknowledge the funding from Heartflow Inc., CA, USA. SCOT-HEART was funded by The Chief Scientist Office of the Scottish Government Health and Social Care Directorates (CZH/4/588), with supplementary awards from Edinburgh and Lothian’s Health Foundation Trust and the Heart Diseases Research Fund.

\end{credits}

%
%
%
%
\bibliographystyle{splncs04}
\bibliography{references}

\newpage
\section{Supplementary}

\subsection{Details of SSA}
Mathematically, when considering the in-plane shifts represented as directions $T_{i,m} = \{\Delta x, \Delta y\}$ for the label map $y_{i}$ of CMR slice $i$ during iteration $m$, the loss function for optimisation is defined as the sum of squared differences (SSD):
\begin{equation}
\label{ssd}
    SSD(T_{i,m}, y_{i,m}) = \sum_{p}(y_{i,m}\circ T_{i,m} - y^{intersect}_{i,m})^{2}
\end{equation}
where $p$ represents the pixel value for the label map $y_{i}$ during iteration $m$. $y^{intersect}_{i,m}$ denotes the intersection with other slices during iteration $m$, and the symbol $\circ$ signifies the application of in-plane shifting to the current label map $y_{i}$. Essentially, the objective of the loss function is to promote the maximisation of the overlapping area between the current label map $y_{i}$ and its intersection with label maps from other slices. 
The pseudocode is presented as follows:\\
\begin{algorithm}
\label{mc_algo}
\SetAlgoLined
{\textbf{Assume} label map $y_{i}$ for each CMR slice, where $i =1,\cdots,N$ and $N$ is the total number of slices.
\text{ }\text{ }\text{ }\text{ }\text{ }\\
 \textbf{For} each iteration $m\in M$:\\ 
  \text{ }\text{ }\text{ }\text{ }\textbf{For} each slice $i\in N$:\\ 
  \text{ }\text{ }\text{ }\text{ }(a) Choose the label map $y_{i}$ from CMR slice (called as moving image).\\
  \text{ }\text{ }\text{ }\text{ }(b) Calculate the intersections between the moving image and all \text{ }\text{ }\text{ }\text{ }\text{ }\text{ }\text{ }\text{ }remaining label maps $y_{j}$, where $j \neq i$.\\
 \text{ }\text{ }\text{ }\text{ }(c) Resample all interactions into one image (called as intersection \text{ }\text{ }\text{ }\text{ }\text{ }\text{ }\text{ }\text{ }image).\\
  \text{ }\text{ }\text{ }\text{ }(d) Maximize the overlapping between moving image and intersection \text{ }\text{ }\text{ }\text{ }\text{ }\text{ }\text{ }\text{ }image with loss given by \ref{ssd}, which outputs in-plane adjustment.\\
  \text{ }\text{ }\text{ }\text{ }(e) Apply the in-plane shift for the current moving image $y_{i}$.\\
  \text{ }\text{ }\text{ }\text{ }\text{ }\\
   \text{ }\text{ }\text{ }\text{ }\text{ }Check the convergence criteria}.
 \caption{Slice Shifting Algorithm (SSA)}
\end{algorithm}
\subsection{Efficiency comparison for SSA}
Under our implementation of SSA in \textit{PyTorch}, the running time of per case is approximately 20 seconds, whereas it takes around 2 minutes for the original implementation \cite{mattmotion}.
\subsection{Visualisation of SSA}
\begin{figure}[H]
    \centering
    \includegraphics[scale=0.25]{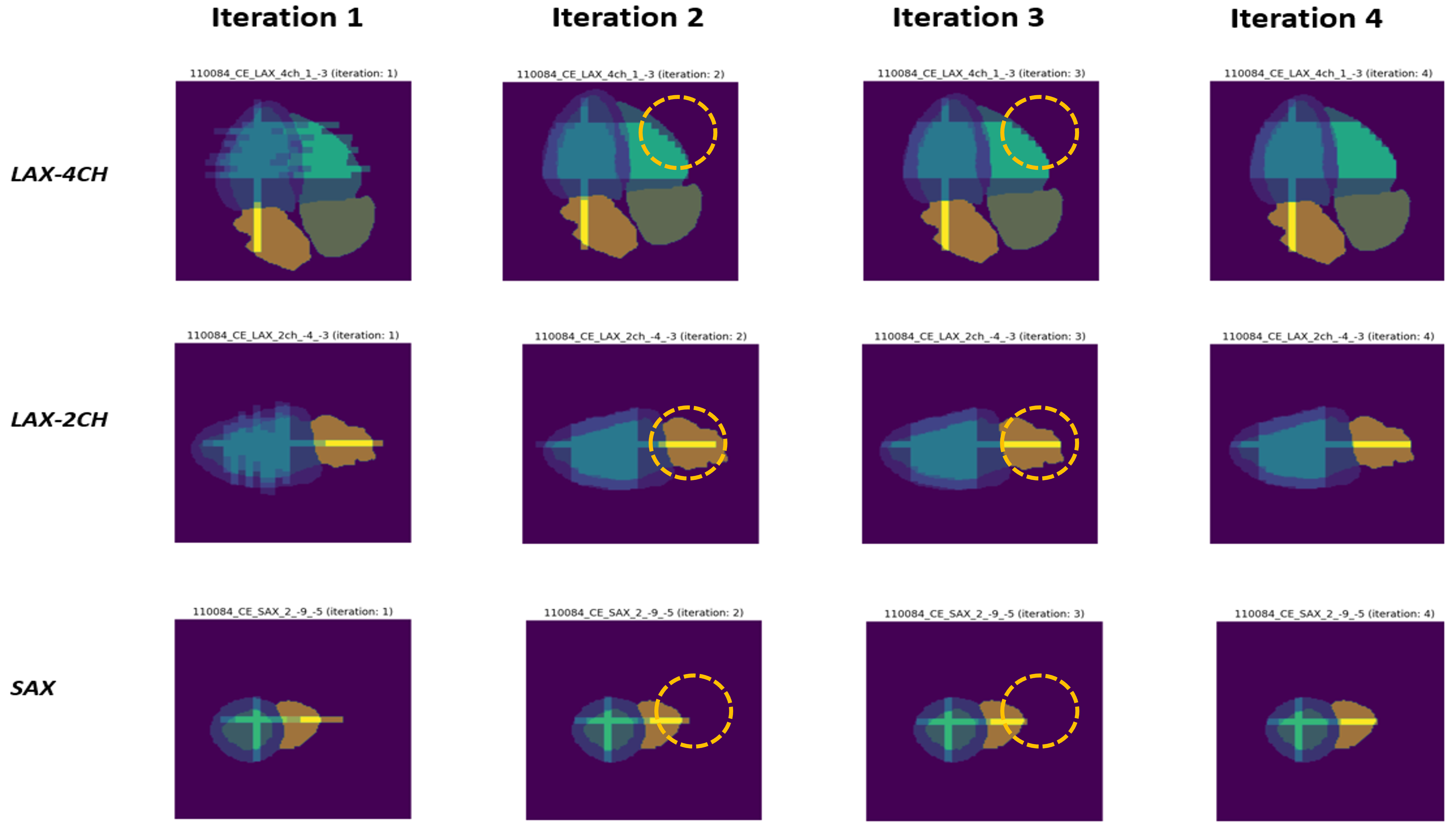}
    \caption{SSA for synthetic CMR data: it shows the intersection between the current slice and other slices under each iteration. Convergence is achieved after 3 iterations.}
\end{figure}

\end{document}